\documentclass[aps,pra,preprint,11pt]{revtex4-1}

\usepackage{amssymb}
\usepackage{amsmath}
\usepackage{bm}
\usepackage{natbib}
\usepackage{graphics,graphicx}
\usepackage{epsfig}
\usepackage{latexsym}
\usepackage{epic,eepic,graphicx,amssymb,amsmath,indentfirst}
\usepackage{bm}
\usepackage{graphicx}
\usepackage[english]{babel}
\usepackage[latin1]{inputenc}
\usepackage{multirow}

\usepackage{color}
\usepackage[colorlinks={true}]{hyperref}
\hypersetup{citecolor={blue}, filecolor={blue}, linkcolor={blue}, urlcolor={blue}}
\newcommand{\bra}[1]{\langle #1 \vert}

\begin{document}

\title{\bf Coherent control of nonlinear mode transitions in Bose-Einstein condensates}

\author{Leonardo Brito da Silva}
\email{brito@if.usp.br}
\affiliation{Instituto de F\'isica, Universidade de S\~ao Paulo\\ S\~ao Paulo, SP 05508-090, Brazil}

\author{Emanuel Fernandes de Lima}
\email{emanuel@df.ufscar.br}
\affiliation{Departamento de F\'isica, Universidade Federal de S\~ao Carlos\\ S\~ao Carlos, SP 13565-905, Brazil}
\date{\today}

\begin{abstract}
 We investigate the formation of non-ground-state Bose-Einstein condensates within the mean-field description represented by the Gross-Pitaevskii equation (GPE). The goal is to form excited states of a condensate known as nonlinear topological modes, which are stationary solutions of the GPE. Nonlinear modes can be generated by modulating either the trapping potential or the atomic scattering length. We show that it is possible to coherently control the transitions to excited nonlinear modes by manipulating the relative phase of the modulations. In addition, we show that the use of both modulations can modify the speed of the transitions. In our analysis, we employ approximate analytical techniques, including a perturbative treatment, and numerical calculations for the GPE. Our study evidences that the coherent control of the GPE presents novel possibilities which are not accessible for the Schr\"odinger equation.
\end{abstract}

\maketitle

\section{Introduction}

The interference between quantum transition amplitudes is the underlying principle of coherent control \cite{doi:10.1146/annurev.pc.43.100192.001353,doi:10.1146/annurev.physchem.48.1.601,doi:10.1146/annurev.physchem.040808.090427}. Given some independent excitation pathways connecting an initial state to a final target state of a quantum system, the total transition probability depends on the modulus squared of the sum of the corresponding complex transition amplitudes. Therefore, upon changing the relative phase of these amplitudes, it is possible to modify the final yield. This concept has found several applications in atomic, molecular and semiconductor physics, among other areas \cite{Petta2180}. In addition, it has contributed towards the development of quantum optimal control theory (QOCT), which seeks to find external controls to drive a given transition, maximizing some performance criteria \cite{rabitz2000,brif2010,peirce1988}. 

Coherent control has been applied in the context of ultracold atomic gases \cite{koch2012,PhysRevLett.96.173002,PhysRevA.75.051401,PhysRevA.95.013411} and in particular to Bose-Einstein condensates (BEC) \cite{Thomasen_2016,Bohi2009,Lozada_Vera_2013}. For instance, the control of the onset of self-trapping of a condensate in a periodically modulated double well has been demonstrated \cite{PhysRevA.64.011601}. Also, SU(2) rotations on spinor condensates has been coherently controlled by stimulated Raman adiabatic passage \cite{PhysRevA.94.053636}. In addition, there have been several experimental and theoretical successful implementations of quantum optimal control algorithms for BEC \cite{Nat1,Nat2,hohenester2014,hohenester2007,Mennemann_2015,0953-4075-46-10-104012,PhysRevA.90.033628,PhysRevA.98.022119}.    

A challenging control problem is the generation of non-ground-state condensates, which can exhibit several interesting features such as mode locking, critical dynamics, interference patterns and atomic squeezing \cite{yukalov69}. In the mean-field picture, non-ground-states can be represented as stationary solutions of the Gross-Pitaevskii equation (GPE), and are termed nonlinear topological modes \cite{yukalov66}. It has been shown that a given nonlinear mode can be generated by a time-dependent modulation of the trapping potential tuned to the resonance between the ground state and the desired excited nonlinear mode \cite{yukalov56}. Alternatively, it is also possible to generate a nonlinear mode by means of a resonant time-dependent modulation of the scattering length \cite{yukalov78,PhysRevA.81.053627}, which can be produced by an alternating magnetic field close to a Feshbach resonance \cite{PhysRevA.67.013605,PhysRevA.81.053627,clark2015,arunkumar2019}. Since quantum transition amplitudes can be attributed to these different modulations, their relative phase can be used to control the overall transition. Nevertheless, due to the nonlinear nature of the GPE, the transition amplitudes of the trapping and scattering length modulations are not completely independent, and new aspects of the coherent control can be expected in comparison with the linear Schr\"odinger equation. It is worth noting that QOCT has been applied to maximize the transition between nonlinear modes utilizing temporal and spatial modulation of both the trapping and the scattering length \cite{,PhysRevA.93.053612}. However, the control fields obtained from QOCT are often complex and difficult to be implemented in laboratory. Additionally, the detailed role of the interference between both modulations has not been addressed. 

In the present work, we explore the possibility of using both modulations to resonantly generate non-ground-state condensate. In particular, we consider the role of the relative phase of the modulations on the transitions between nonlinear modes. In sec.~\ref{sec2}, we introduce the theoretical framework for the production of the nonlinear coherent modes. In sec.~\ref{twolevels}, a time average technique is applied to the GPE allowing the description of the dynamics by only two modes. A perturbative treatment is used to obtain an analytic expression for the transition probability in sec.~\ref{sec4}. Numerical results confirming the predictions are presented in sec.~\ref{sec5}. Finally, conclusions are drawn in sec.~\ref{sec6}.

\section{Excitation of nonlinear modes}\label{sec2}

We consider the dynamics of a Bose gas wavefunction $\Psi({\bf r},t)$ described by the Gross-Pitaevskii equation \cite{RevModPhys.71.463,pethick},
\begin{equation}
{\rm i}\hbar\frac{\partial}{\partial t}\Psi({\bf r},t)=\left(-\frac{\hbar^2}{2m}\nabla^2+V({\bf r},t)+g({\bf r},t)N|\Psi({\bf r},t)|^2\right)\Psi({\bf r},t),
\label{eq:TDGPE}
\end{equation}
where $m$ is the atomic mass and $N$ is the number of atoms in the condensate. The trapping potential $V({\bf r},t)$ is composed by two parts,
\begin{equation}
V({\bf r},t)=V_{\rm trap}({\bf r})+V_{\rm mod}({\bf r},t),
\label{eq:trappingwm}
\end{equation}
where $V_{\rm trap}({\bf r})$ is a fixed trapping potential and $V_{\rm mod}({\bf r},t)$ a time-dependent modulating potential. The nonlinear interaction amplitude $g({\bf r},t)$ is also composed by two parts,
\begin{equation}
g({\bf r},t)=g_{0}+g_{\rm mod}({\bf r},t),
\label{eq:nonlinearitygwm}
\end{equation}
where $g_0=4\pi\hbar^2a_0/m$ is a fixed nonlinearity and $g_{\rm mod}({\bf r},t)=4\pi\hbar^2a({\bf r},t)/m$ is a modulating nonlinearity, with the s-wave scattering length $a_s$ near a Feshbach resonance being written as $a_s=a_0+a({\bf r},t)$. The normalization condition of the wavefunction is $\int d{\rm \bf r}|\Psi({\bf r},t)|^2=1$.

With both modulations turned off, i.e., $V_{\rm mod}({\bf r},t)=0$ and $g_{\rm mod}({\bf r},t)=0$, the system can be described by the nonlinear Hamiltonian $H_0$,

\begin{equation}
    H_0[\phi({\bf r})]=-\frac{\hbar^2}{2m}\nabla^2+V_{\rm trap}({\bf r})+{g}_{0}N|\phi({\bf r})|^2.
\end{equation}

We consider the nonlinear topological modes of $H_0$, which are solutions of the eigenvalue problem \cite{yukalov56},
\begin{equation}
    H_0[\phi_n({\bf r})]\phi_n({\bf r})=\mu_n\phi_n({\bf r}),
    \label{multiindex0}
\end{equation}
with $n$ generally being a multi-index label for the quantum states and $\mu_n$ the corresponding chemical potential. Here, we are concerned with inducing transitions between stationary solutions $\phi_n({\bf r})$. This task can be accomplished by means of modulating the trapping potential with an oscillatory field with frequency $\omega_t$ \cite{yukalov56,courteille}. Alternatively, one may also modulate the atomic scattering length with frequency $\omega_g$ \cite{yukalov78}. We assume that both modulations are present and that they possess a phase difference given by $\theta$,

\begin{equation}
    V_{\rm mod}({\bf r},t)=V({\bf r})\cos(\omega_t t+\theta),
\label{Vmod}
\end{equation}
and 
\begin{equation}
   g_{\rm mod}({\bf r},t)=g({\bf r})\cos(\omega_g t).
   \label{gmod}
\end{equation}

For definiteness, we consider a transition between an initial state $\phi_1({\bf r})$ to a final state $\phi_2({\bf r})$, with $\mu_1<\mu_2$, and we associate the resonance frequency $\omega_{21}=(\mu_2-\mu_1)/\hbar$ with this transition. As we have already pointed out, the transition can be induced by resonant modulations and in this case the system can be approximately described solely by the topological modes involved in the transition \cite{yukalov56,courteille}. Thus, we assume that the modulating frequencies $\omega_t$ and $\omega_g$ are close to $\omega_{21}$. More specifically, we assume that $|\Delta\omega_{t}/\omega_t|\ll1$ and $|\Delta\omega_{g}/\omega_g|\ll1$, with the detunings defined by $\Delta\omega_{t}=\omega_{t}-\omega_{21}$ and $\Delta\omega_{g}=\omega_{g}-\omega_{21}$.

\section{Two-Level approximation}\label{twolevels}

In order to simplify the dynamical equations, we consider that the wave function $\Psi({\bf r},t)$ can be written as an expansion in terms of nonlinear modes \cite{yukalov56}, 
\begin{equation}
\Psi( {\bf r},t) =\sum_{m}c_{m}(t)\phi_{m}({\bf r})\exp\left(-i\mu_m t/\hbar\right),
\label{expan0}
\end{equation}
and that the following condition for mode separation is valid,
\begin{equation}
    \frac{\hbar}{\mu_m} \left|\frac{dc_{m}}{dt}\right|\ll1,
    \label{condvary0}
\end{equation}
meaning that the $c_m(t)$ are slow function of time in comparison with $\exp(-i\mu_m t/\hbar)$.

Substituting expansion (\ref{expan0}) in the GPE and performing a time-averaging procedure, with the coefficients $c_m(t)$ treated as quasi-invariants, will result in a set of coupled nonlinear differential equations for the coefficients $c_m(t)$ (see Ref. \cite{yukalov66} for details). As a consequence, if at the initial time only  the levels $n=1,2$ are populated and the frequencies of the modulations are close to $\omega_{21}$, the only relevant coefficients are $c_1(t)$ and $c_2(t)$ and we obtain the set of equations,

%\begin{align}
\begin{subequations}\label{pop2}
\begin{alignat}{2}
i\frac{d{c}_{1}}{dt} = &  {\alpha}_{12}{\left| c_{2} \right|}^{2}{c}_{1} +\frac{1}{2}{\beta}_{12}c_{2}{\rm{e}}^{i\left(\Delta \omega_{t} t +\theta \right)}+ \frac{1}{2}{\rm{e}}^{-i\Delta\omega_{g} t}c_{2}^{*}{c_{1}}^{2}\gamma_{21}  \nonumber
 \\ & +\frac{1}{2}{\rm{e}}^{i\Delta\omega_{g} t}\left( {\left|c_{2} \right|}^{2}c_{2}\gamma_{12} + 2{\left|c_{1} \right|}^{2}c_{2}{\gamma}_{21}^{*} \right)   \ ,
 \\
i\frac{d{c}_{2}}{dt}=  &  {\alpha}_{21}{\left| c_{1} \right|}^{2}{c}_{2}+\frac{1}{2}{\beta}_{12}^{*}c_{1}{\rm{e}}^{-i\left(\Delta \omega_{t} t +\theta\right)} +
\frac{1}{2}{\rm{e}}^{i\Delta\omega_{g} t}c_{1}^{*}{c_{2}}^{2}\gamma_{12} \nonumber
 \\ & + \frac{1}{2}{\rm{e}}^{-i\Delta\omega_{g} t}\left( {\left|c_{1} \right|}^{2}c_{1}\gamma_{21} + 2{\left|c_{2} \right|}^{2}c_{1}{\gamma}_{12}^{*} \right),
  \end{alignat}
 \end{subequations}
with the coupling constants $\alpha_{mn}$, $\beta_{mn}$ and $\gamma_{mn}$ given by
\begin{equation}
    \alpha_{mn} \equiv g_{0}\frac{N}{\hbar}\int{d{\bf r}{|\phi_{m}({\bf r})|}^{2}\left[ 2{|\phi_{n}({\bf r})|}^{2}-{|\phi_{m}({\bf r})|}^{2}\right]},
    \label{alpha0}
\end{equation}
\begin{equation}
    \beta_{mn}\equiv \frac{1}{\hbar}\int{d{\bf r}{\phi}_{m}^{*}({\bf r})V({\bf r}){\phi}_{n}({\bf r})} \ ,
    \label{beta0}
\end{equation}
and
\begin{equation}
    \gamma_{mn}\equiv \frac{N}{\hbar}\int{d{\bf r}{\phi}_{m}^{*}({\bf r})g({\bf r}){|\phi_{n}({\bf r})|}^{2}}\phi_{n}({\bf r}) \ .
    \label{I0}
\end{equation}
In order to fulfill condition (\ref{condvary0}), the couplings are assumed to be much smaller than the transition frequency, i.e., $ |\alpha_{mn}/\omega_{21}|\ll1 $,  $            |\beta_{mn}/\omega_{21} |\ll1 $ and $ |\gamma_{mn}/\omega_{21} |\ll1 $ \cite{yukalov66}.

From the dynamical equations (\ref{pop2}), we note that the modulation of the trap couples the modes by means of the linear term containing $\beta_{mn}$ and the nonlinear term with $\alpha_{mn}$, while the modulation of the scattering length couples the modes by means of distinct nonlinear terms containing $\gamma_{mn}$. Therefore, there exists interference between the linear and nonlinear terms and by varying the phase $\theta$, this interference can be controlled. We also note that when the modulation of the scattering length is absent, $g_{\rm mod}({\bf r},t)=0$, an approximate analytic solution to (\ref{pop2}) has been derived, which shows that the population oscillates between the two states with a Rabi-like chirped frequency \cite{yukalov56}. This chirped frequency depends on the populations $|c_m(t)|^2$. Unfortunately, such approximate solutions for $g_{\rm mod}({\bf r},t)\neq0$ is not possible due to the presence of terms with $c_m(t)^2$. Thus, we resort to perturbation theory to gain more insight into the role of $\theta$ in the transition.  

\section{Perturbative Approximation}\label{sec4}

We assume that modulating fields (\ref{Vmod}) and (\ref{gmod}) can be considered as small perturbations in order to apply canonical perturbation theory \cite{cohen}. To this end, we introduce a perturbation parameter $\lambda\ll1$ such that the Hamiltonian can be written as

\begin{equation}
    H[\Psi]=H_0[\Psi]+\lambda \left[V_{\rm mod}({\bf r},t)+g_{\rm mod}({\bf r},t)N|\Psi|^2\right] \ .
    \label{pertH}
\end{equation}
We are interested in the transition probability from state $\phi_1$ to state $\phi_2$, often defined as $P_{1\rightarrow 2}(t)=\left|\bra{\phi_2}\left.\Psi({\bf r},t)\right>\right|^2$ and we assume the initial conditions $c_1(0)=1$ and $c_2(0)=0$. However, from the approximations of the last section, one can deduce the normalization condition for the coefficients, $\sum_{m}\left|c_m(t)\right|^2=1$. Thus, despite of the fact that the set of nonlinear modes is not orthogonal, we can define the transition probability simply as $P_{1\rightarrow 2}(t)=\left|c_2(t)\right|^2$.  

As usual, we write the coefficients $c_{j}(t)$ as a power series in $\lambda$,
\begin{equation}
    c_{j}(t)={c}_{j}^{(0)}(t)+\lambda{c}_{j}^{(1)}(t)+{\lambda}^{2}{c}_{j}^{(2)}(t)+\cdots,
    \label{cserie}
\end{equation}
and substitute this series into dynamic equations (\ref{pop2}) equating the like powers of $\lambda$. To zeroth order, this yields,
\begin{subequations}\label{zerotho}
\begin{alignat}{2}
i\frac{d{c}_{1}^{(0)}}{dt}=\alpha_{12}{|{c}_{2}^{(0)} |}^{2}{c}_{1}^{(0)} ,
\label{c1lambda0} \\
i\frac{d{c}_{2}^{(0)}}{dt}=\alpha_{21}{|{c}_{1}^{(0)}|}^{2}{c}_{2}^{(0)} \ .
\label{c2lambda0}
  \end{alignat}
 \end{subequations}
And we obtain the zeroth order solution as being ${c}_{1}^{(0)}(t)=1$, ${c}_{2}^{(0)}(t)=0$.

To first order in $\lambda$, the equations are 
\begin{subequations}\label{firsto}
\begin{alignat}{2}
\label{c1lambda1} i\frac{d{c}_{1}^{(1)}}{dt}=\alpha_{12}\left[\left({c}_{2}^{*(0)}{c}_{2}^{(1)}+{c}_{2}^{*(1)}{c}_{2}^{(0)} \right){c}_{1}^{(0)} +{|{c}_{2}^{(0)} |}^{2}{c}_{1}^{(1)} \right]+\frac{1}{2}{\beta}_{12}{c}_{2}^{(0)}{\rm{e}}^{i\left(\Delta\omega_{t} t+\theta \right)} \nonumber \\
+\frac{1}{2}{\rm{e}}^{i\Delta\omega_{g} t}\left[ {| {c}_{2}^{(0)}|}^{2}{c}_{2}^{(0)}\gamma_{12}+2{|{c}_{1}^{(0)} |}^{2}{c}_{2}^{(0)}{\gamma}_{21}^{*} \right]+\frac{1}{2}{\rm{e}}^{-i\Delta\omega_{g} t} {c}_{2}^{*(0)}{{c}_{1}^{(0)}}^{2}\gamma_{21} \ , \\
 i\frac{d{c}_{2}^{(1)}}{dt}=\alpha_{21}\left[\left({c}_{1}^{*(0)}{c}_{1}^{(1)}+{c}_{1}^{*(1)}{c}_{1}^{(0)} \right){c}_{2}^{(0)} +{|{c}_{1}^{(0)} |}^{2}{c}_{2}^{(1)} \right]+\frac{1}{2}{\beta}_{12}^{*}{c}_{1}^{(0)}{\rm{e}}^{-i\left(\Delta\omega_{t} t+\theta \right)} \nonumber \\
+\frac{1}{2}{\rm{e}}^{-i\Delta\omega_{g} t}\left[ {| {c}_{1}^{(0)}|}^{2}{c}_{1}^{(0)}\gamma_{21}+2{|{c}_{2}^{(0)} |}^{2}{c}_{1}^{(0)}{\gamma}_{12}^{*} \right]+\frac{1}{2}{\rm{e}}^{i\Delta\omega_{g} t} {c}_{1}^{*(0)}{{c}_{2}^{(0)}}^{2}\gamma_{12} \ .
\label{c2lambda1}
  \end{alignat}
 \end{subequations}

Substituting the zeroth order solutions into (\ref{c1lambda1}) and (\ref{c2lambda1}), these equations simplify to

\begin{subequations}\label{firsto2}
\begin{alignat}{2}
i\frac{d{c}_{1}^{(1)}}{dt} & =0 \ ,
\label{c11lambda1} \\
i\frac{d{c}_{2}^{(1)}}{dt} & =\alpha_{21}{|{c}_{1}^{(0)}|}^{2}{c}_{2}^{(1)}+\frac{1}{2}{\beta}_{12}^{*}{\rm{e}}^{-i\left(\Delta\omega_{t} t +\theta \right)}+\frac{1}{2}\gamma_{21}{\rm{e}}^{-i\Delta\omega_{g} t} \ .
\label{c22lambda1}
  \end{alignat}
 \end{subequations}
Thus, within first order, ${c}_{1}^{(1)}(t)=0$ and 
\begin{equation}
   {c}_{2}^{(1)}(t)=-\frac{1}{2}\frac{{\beta}_{12}^{*}}{(\alpha_{21}-\Delta\omega_{t})}{\rm{e}}^{-i\theta}\left( {\rm{e}}^{-i\Delta\omega_{t} t}-1\right)-\frac{1}{2}\frac{\gamma_{21}}{(\alpha_{21}-\Delta\omega_{g})}\left( {\rm{e}}^{-i\Delta\omega_{g}t}-1\right) \ .
    \label{sc22lambda1}
\end{equation}
Thus, we can write the transition probability within first order as
\begin{eqnarray}
P_{1\rightarrow 2}(t)\approx \frac{{|\beta_{12}|}^{2}}{2{|\alpha_{21}-\Delta\omega_{t}|}^{2}}\left[1- \cos(\Delta\omega_{t} t)\right]+\frac{{|\gamma_{21}|}^{2}}{2{|\alpha_{21}-\Delta\omega_{g}|}^{2}}\left[1- \cos(\Delta\omega_{g} t)\right] \nonumber \\
+ \frac{{\beta}_{12}\gamma_{21}}{4{{(\alpha_{21}-\Delta\omega}_{t})}^{*}(\alpha_{21}\Delta\omega_{g})}{\rm{e}}^{i\theta}\left[1+{\rm{e}}^{i(\Delta\omega_{t}-\Delta\omega_{g})t}-{\rm{e}}^{i\Delta\omega_{t}t}-{\rm{e}}^{-i\Delta\omega_{g}t}\right]  \nonumber \\
+\frac{{\gamma}_{21}^{*}{\beta}_{12}^{*}}{4{(\alpha_{21}-\Delta{\omega}_{g})}^{*}(\alpha_{21}-\Delta\omega_{t})}{\rm{e}}^{-i\theta}\left[1+{\rm{e}}^{-i(\Delta\omega_{t}-\Delta\omega_{g})t}-{\rm{e}}^{-i\Delta\omega_{t}t}-{\rm{e}}^{i\Delta\omega_{g}t}\right]    \ .
\label{P12t}
\end{eqnarray}
When the frequencies of the modulations are equal, $\Delta \omega_{t}=\Delta\omega_{g}=\Delta\omega$, the expression for the transition probability simplifies to

\begin{equation}
P_{1\rightarrow 2}\approx \left[  \frac{{|\beta_{12}|}^{2}+{|\gamma_{21}|}^{2}+2\Re\{{\beta}_{12}{\gamma}_{21}\exp(i\theta)\}}{{|\alpha_{21}-\Delta\omega|}^{2}}\right]{\sin}^{2}\left(\frac{\Delta\omega t}{2}\right) \ ,
    \label{P12twr}
\end{equation}
where $\Re\{\cdot\}$ stands for the real part.

Although the above expression is only valid for very short times, for which the population of the state $\phi_1$ is still very close to $1$, expression (\ref{P12twr}) evidences the role of the relative phase $\theta$ on the transition. For instance, if $({\beta}_{12}\gamma_{21})$ is real and positive, then for $\theta=\pi$ the modulations will act destructively decreasing the transition probability, whereas for $\theta=0$ they will act constructively. The extent of the interference will be dictated by the magnitude of the couplings parameters $\beta_{12}$ and $\gamma_{12}$. Additionally, according to (\ref{P12twr}), if the modulation of the scattering length is absent, then variation of $\theta$ plays no role in the dynamics.

\section{numerical results}\label{sec5}

We have carried out direct numerical calculations of the GPE solving Eq.~(\ref{eq:TDGPE}) in its 1D version,

\begin{equation}
{\rm i}\frac{\partial}{\partial t}\Psi(x,t)=H[\Psi]\Psi(x,t),
    \label{eq:TDGPE_1D}
\end{equation}
with the nonlinear Hamiltonian given by
\begin{equation}
 H[\Psi]= -\frac{\partial^2}{\partial x^2}+V(x,t)+g(x,t)|\Psi(x,t)|^2,  
\end{equation}
and considering arbitrary units such that $\hbar=m=N=g_0=1$. The nonlinear Hamiltonian operator has been written as a matrix over a grid of points according to the Chebyshev spectral method \cite{trefethen,mason}.

In order to solve the time-dependent equation (\ref{eq:TDGPE_1D}), we express the corresponding time evolution operator, which connects the initial time $t=0$ to the final time $t=t_f$, in $N$ small time-step $\Delta t$ evolution operators,
\begin{equation}
U(t_f,0)=\prod_{k=1}^{N}U\left(k\Delta t,(k-1)\Delta t\right).
    \label{evolEP}
\end{equation}
Each one of the small time-step evolution operators is calculated as an expansion in Chebyshev polynomials \cite{kosloff1984,LEFORESTIER199159,formanek2010},

\begin{equation}
   U\left(k\Delta t,(k-1)\Delta t\right)\approx\sum_{n=0}^{N_p}a_n\chi_n(-iH[\Psi((k-1)\Delta t)]\Delta t), 
\end{equation}
where $a_n$ are the expansion coefficients, $\chi_n$ are the complex Chebyshev polynomials and $N_p$ sets the number of terms in the expansion. The propagation of the wavefunction in the $k$-th time step is obtained by applying $U\left(k\Delta t,(k-1)\Delta t\right)$ to the wavefunction calculated in the previous step $\Psi((k-1)\Delta t)$. The relaxation method, which in essence consists in performing propagation with imaginary time $t\rightarrow {\rm i}t$, has been applied to obtain the ground state \cite{kosloff1986}. The excited modes of the condensate have been determined by the spectrum-adapted scheme described in Ref. \cite{PhysRevA.93.053612}. We have found very good agreement comparing our results with  those from Refs. \cite{MURUGANANDAM20091888,PhysRevA.93.053612}.

For harmonic trapping potentials and modulating fields with linear behavior with distance, no transition to excited modes is possible through modulation of the trap \cite{yukalov69}. Thus, we have fixed the trapping potential to $V_{trap}(x,t)=x^4/4$, allowing for a simple form of the spatial dependence of $V(x)$. For this trap, we have obtained the chemical potentials $\mu_{0}= 0.808$,  $\mu_{1}=1.857$, and $\mu_{2}= 3.279$, for the ground, first and second nonlinear modes, respectively.

We have considered transitions from the ground state to the first and to the second excited modes. In the first case, we have set $g(x)=A_g x$ and $V(x)=A_t x$, while in the second case, $g(x)=A_g x^2$ and $V(x)=A_t x^2$. The frequencies of the modulations are set to be equal $\omega_t=\omega_g=\omega$ and are chosen to satisfy the resonance condition for each target.

\begin{figure}[ht]
		\begin{center}
					\includegraphics[width=\textwidth]{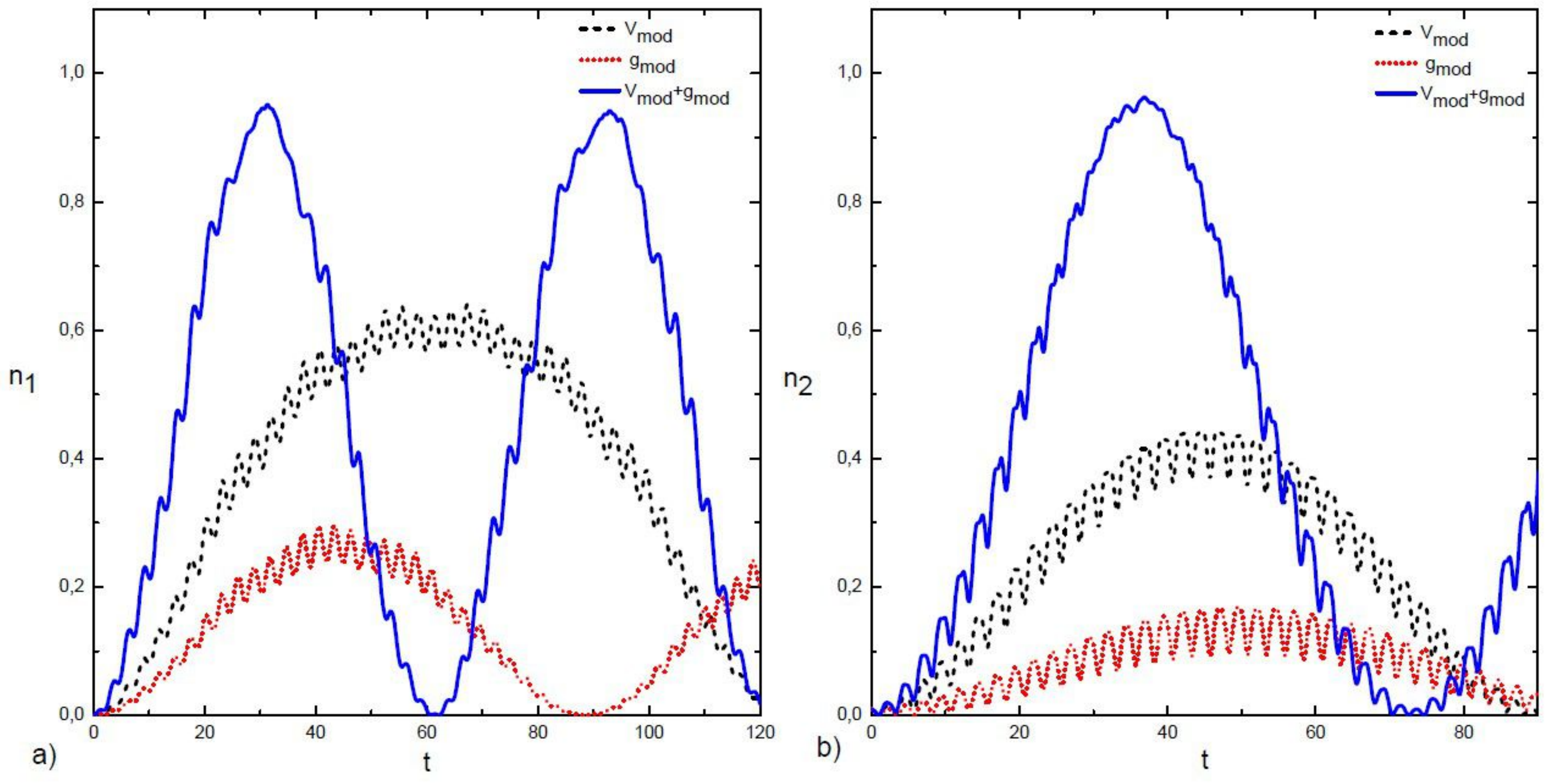}
\end{center}
\caption{\label{figure1} a) Population of first excited state versus time for a system driven by trap, scattering length and double modulation ($\theta=0$), with $A_{t}=0.1$ and $A_{g}=0.3$ amplitudes. b) Population of second excited state by trap, scattering length and double modulations, with $A_{t}=0.08$ and $A_{g}=0.4$ amplitudes.}
\end{figure}

Figure~\ref{figure1} compares single modulation with double modulation for $\theta=0$ by showing the corresponding target population dynamics, denoted by $n_j\equiv\left|\bra{\phi_j}\left.\Psi(x,t)\right>\right|^2$. In panel (a), the target is the first excited state, while in panel (b) the target is the second excited state. In both cases, we observe the double modulation performing a faster transition than the individual modulations. Additionally, the double modulation enhances the target population beyond that of the sum of the individual modulations, which is an evidence of quantum interference.

\begin{figure}[ht]
\begin{center}
	\includegraphics[width=\textwidth]{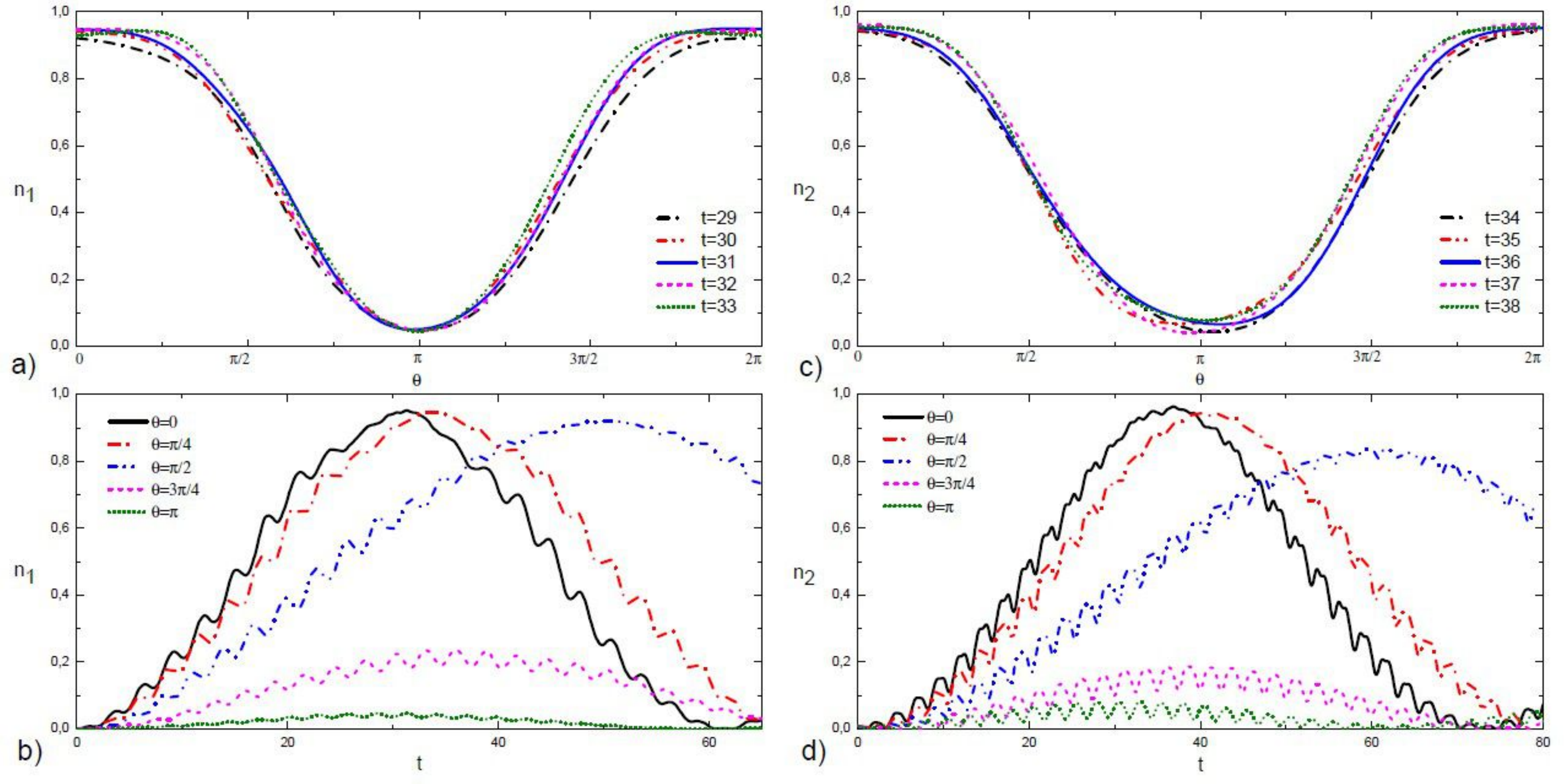}
\end{center}
\caption{\label{figure2} a) Population of the first excited mode versus relative phase of the modulations for some fixed times and parameters of panel a) of Fig. \ref{figure1}. b) Population of the first excited state versus time for some fixed phases (same parameters of a)). c) Population of the second excited state for some fixed times and parameters of panel b) of Fig. \ref{figure1}. d)  of the second excited mode versus time for some fixed phases (same parameters of c)).}
\end{figure}

Panels (a) and (c) of Fig.~\ref{figure2} show the population of the target modes, the first and second modes, respectively, as a function of the relative phase of the modulations for some fixed times. For $\theta=\pi$ the transition is essentially inhibited, whereas for $\theta=0$ the target population is enhanced, in agreement with the perturbative analysis. Panels (b) and (d) show the corresponding population dynamics for some fixed phases. We observe that as the phase varies from $0$ to $\pi$, the transitions become slower while transferring fewer atoms. This behavior has not been captured by the perturbative expression and may be attributed to the nonlinear character of the GPE.

\begin{figure}[ht]
	\begin{center}
		\includegraphics[width=\textwidth]{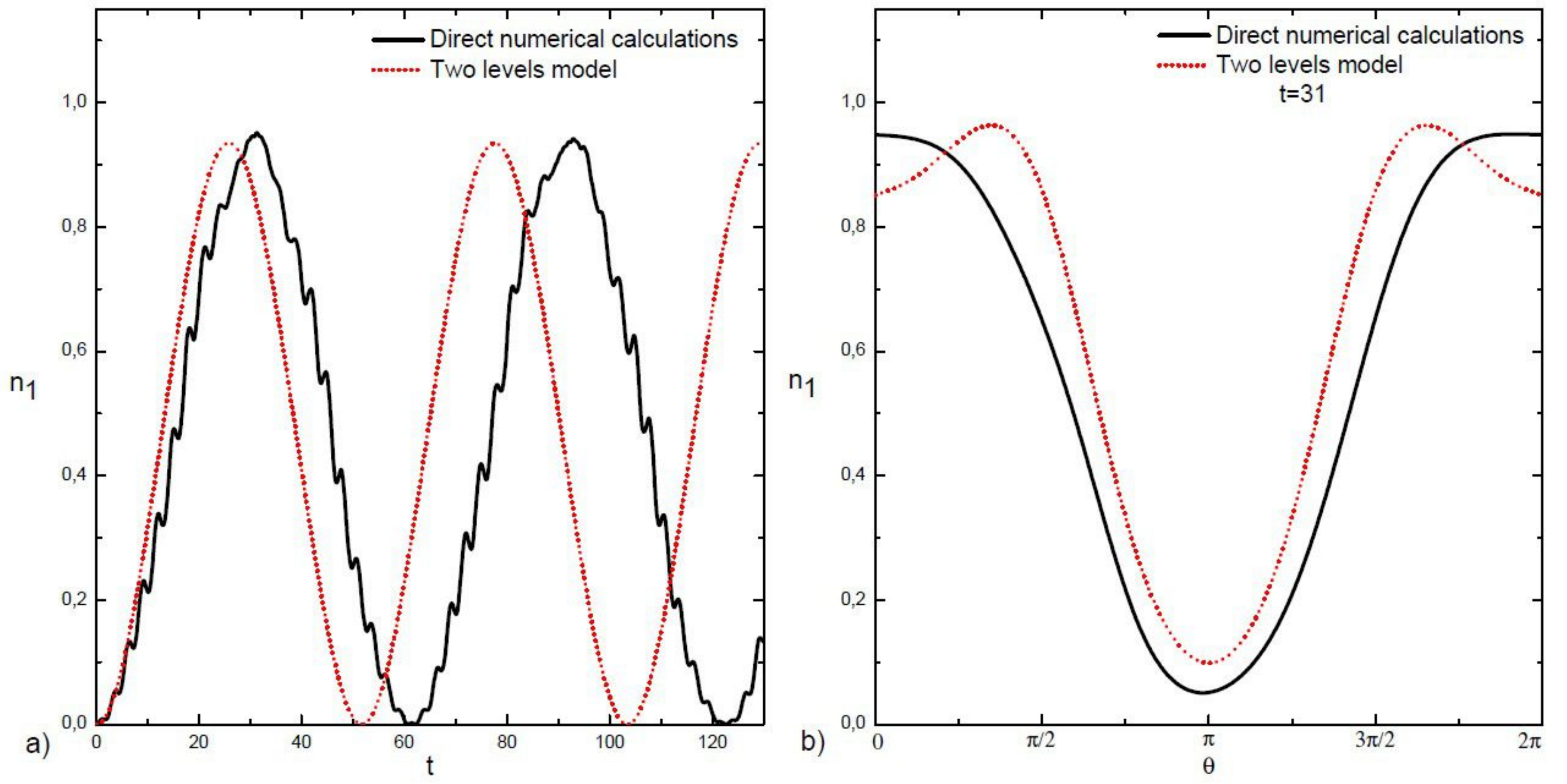}
\end{center}
\caption{\label{figure3} a) Population of the first excited mode versus time under comparison of the direct numerical calculations and the two levels model for the same system and parameters as we introduced in Fig.  \ref{figure1} with $\theta=0$ phase . b) Population versus relative phase of the modulations at $t=31$.}
\end{figure}

Figure~\ref{figure3} illustrates the comparison of the two-level approximation obtained by solving Eq.~(\ref{pop2}) using a fourth-order Runge-Kutta method with the numerical solution of the GPE. We have obtained for the coupling parameters: $\alpha_{21}\approx0.124$, $\beta_{12}\approx7.7\times{10}^{-2}$ and $\gamma_{21}\approx4.6\times{10}^{-2}$ for the transition to the first excited mode. Panel (a) shows the target population dynamics for $\theta=0$, while panel (b) shows the target population at $t=31$ as a function of the relative phase. Although the two-level approximation departs from the expected solution as the evolution takes place, these two panels illustrate the good qualitative agreement between the approaches that we have generally found in our calculations, corroborating our analysis. 

\begin{figure}[ht]
		\begin{center}
	\includegraphics[width=\textwidth]{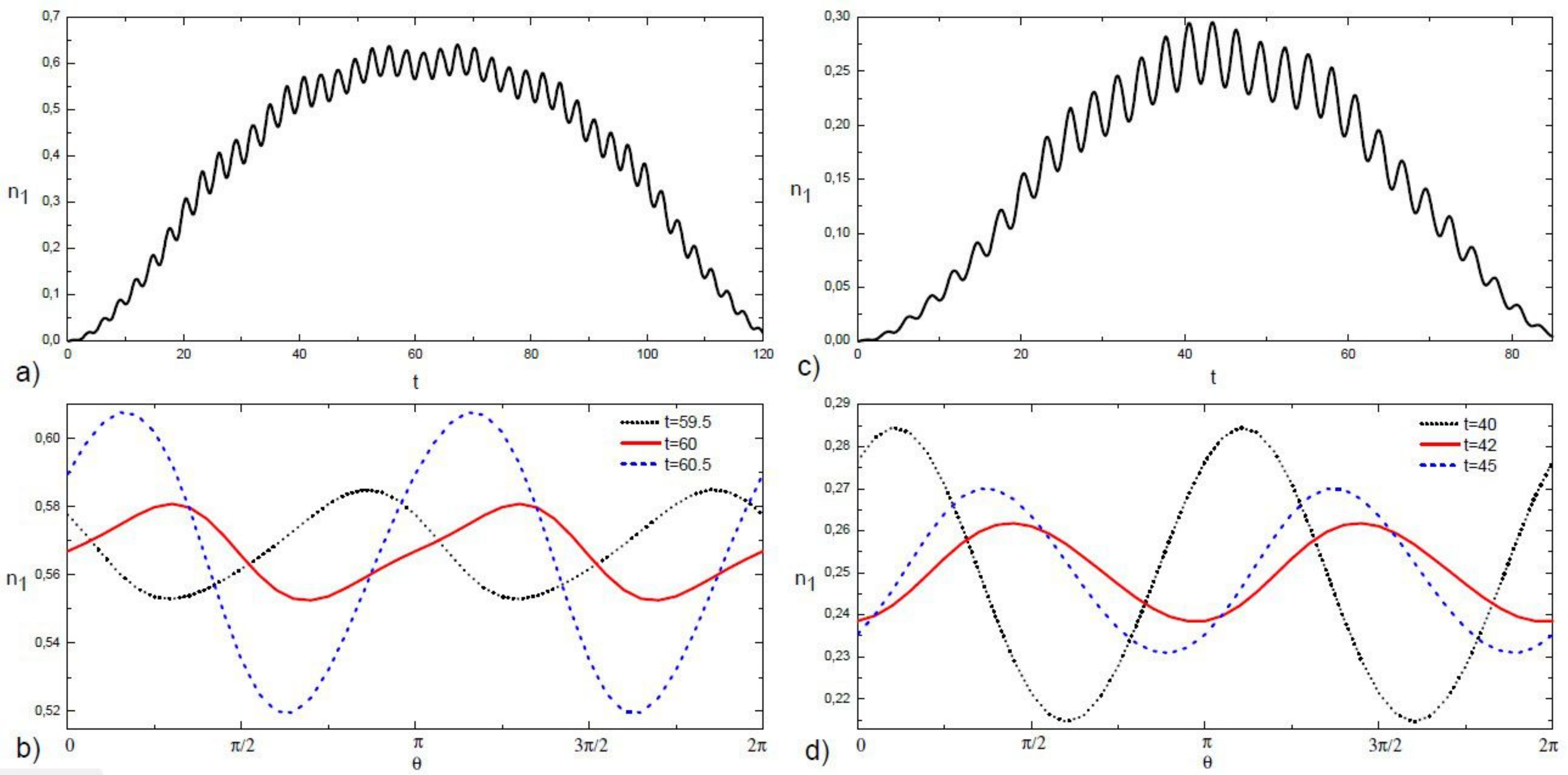}
\end{center}
\caption{\label{figure4} a) Population of first excited mode versus time for a system driven by the trap modulation only, with $\theta=0$ and $A_{t}=0.1$ amplitude. b) Population versus relative phase for some fixed times (same parameters as a)). c) Population of first excited state versus time for a system driven by the scattering length modulation only with $\theta=0$ and $A_{g}=0.3$ amplitude. d) Population of the first excited mode versus relative phase for some fixed times (same parameters as c).}
\end{figure}

Figure \ref{figure4} considers the impact of the phase when only a single modulating field is present. In panels (a) and (b), the modulation of the nonlinearity is turned off $g_{\rm mod}(x,t)=0$, whereas in panels (c) and (d) the modulation of the trap is turned off $V_{\rm trap}(x,t)=0$ and Eq.(\ref{gmod}) reads $g_{\rm mod}=g(x)\cos(\omega_g t+\theta)$. The upper panels show the population dynamics of the first mode, while the lower panels show the population of the first mode as a function of $\theta$ for some fixed times. In contrast with the prediction of the perturbative approach, in both cases the phase does have an impact on the population dynamics. However, this impact is indeed very small compared to the case when the two modulating fields are present.

\section{Conclusion}\label{sec6}

Non-ground-state BEC can be created from the ground state by resonantly modulating either the trapping potential or the atomic interactions. We have explored the simultaneous use of the modulations on the population dynamic and the impact of their relative phase on the formation of the excited modes in the framework of the GPE. Numerical as well as approximated analytical methods have been applied. We have shown that the relative phase can be used to coherently control the transition to the excited modes by enhancing or suppressing the transition. We have also shown that the double modulation can affect the speed of the transitions. This behavior, which is not often found in ordinary quantum dynamics, can be attributed to the nonlinear character of the GPE. This work should stimulate the search for experimental evidences of coherent control of transitions between nonlinear modes induced by double modulation. It should also motivate the study of different control problems using double modulation.

\section*{Acknowledgments}\label{sec7}

This study was financed in part by the Coordination for the Improvement of Higher Education Personnel (CAPES) - Finance Code 001. EFL acknowledges support from Sao Paulo Research Foundation, FAPESP (grant 2014/23648-9) and from National Council for Scientific and Technological Development, CNPq (grant 423982/2018-4).

\pagebreak

\pagebreak

\pagebreak

\pagebreak

\end{document}